# Anomalous behavior of the Hall effect in electron-doped superconductor $Nd_{2-x}Ce_xCuO_{4+\delta}$ with nonstoichiometric disorder

T.B.Charikova[1,*], N.G.Shelushinina[1], G.I.Harus[1], V.N.Neverov[1], D.S.Petukhov[1], O.E.Sochinskaya[1], A.A.Ivanov[2]

[1]*Institute of Metal Physics RAS, Ekaterinburg, Russia,*
[2]*Moscow Engineering Physics Institute, Moscow, Russia*

[*]e-mail: charikova@imp.uran.ru

**Abstract.** Magnetoresistivity and Hall effect measured in magnetic fields up to $B$=9T ($B \parallel c$, $J \parallel ab$) in electron-doped $Nd_{2-x}Ce_xCuO_{4+\delta}$ single crystal films with $x$ = 0.14; 0.15; 0.18 and different oxygen content ($\delta$) were studied in a temperature range of 0.4-4.2 K. The resistivity and Hall coefficient behaviors in the mixed state are discussed in the framework of flux-flow model with the inclusion of the back-flow of vortices owing to the pinning forces.

**Introduction**

In the mixed state of type-II superconductor the onset of dissipation takes place due to hydrodynamical forces acting on moving vortices. In presence of transport current $J$, vortices move in response to a Lorentz force $F = J \times B$ with a velocity $v_L$. As a consequence an electric field $E = -v_L \times B$ is induced which has resistive component $E_x$ and Hall component $E_y$. the longitudinal resistivity, $\rho_{xx} = E_x / J$, and Hall resistivity, $\rho_{xy} = E_y / J$, thus appear.

The Hall effect in the mixed state has been an important problem for the understanding of flux motion in type-II superconductors. An anomalous sign reversal of $\rho_{xy}$ in the mixed state with respect of its normal state often observed both in low-$T_c$ and hole-doped high-$T_c$ type-II superconductors [1]. This effect was in contradiction with existing theories for the flux-vortex motion in perfectly homogeneous structures (flux-flow models [2, 3]).

Wang and Ting [4] demonstrated that the observed negative Hall resistivity in mixed state of certain hole-doped high-$T_c$ superconductors can be explained in terms of pinning forces existing in the sample. In electron-doped high-$T_c$ superconductors mixed state Hall anomaly was observed earlier for some samples by Hagen et al. in 1993 ($Nd_{2-x}Ce_xCuO_{4+\delta}$ single crystal and films with $x$ = 0.15) [5], by Cagigal et al. in 1994 ($L_{1.85}Ce_{0.15}CuO_{4+\delta}$ single crystals; L=Nd, Sm) [6] and by Harus et al in 1997 ($Nd_{2-x}Ce_xCuO_{4+\delta}$ single crystal films with $x$ = 0.18) [7].

In the electron-doped superconductors $Nd_{2-x}Ce_xCuO_{4+\delta}$ superconductivity depends not only on the number of doped carriers putting into the cooper oxygen planes but on the additional oxygen reduction. Oxygen affects the critical temperature and also the upper critical field, the resistivity and the Hall effect. In this paper, we study the effect of nonstoichiometric disorder on the magnetoresistivity and Hall coefficient in mixed and normal states in underdoped ($x$=0.14), optimally doped ($x$=0.15) and overdoped ($x$=0.18) regions of electron doped single crystal films $Nd_{2-x}Ce_xCuO_{4+\delta}$ with different oxygen content $\delta$ (the degree of disorder) in magnetic fields up to 9T ($B \parallel c$, $J \parallel ab$) and in the temperature range of 0.4-4.2K.

**Experiment**

The epitaxial $Nd_{2-x}Ce_xCuO_{4+\delta}$ films ($x$ = 0.14, 0.15, 0.18) with the (001) orientation, where the $c$ axis is perpendicular to the $SrTiO_3$ substrate, were synthesized at the Moscow Engineering Physics Institute using the pulsed laser sputtering method [8]. The films were subjected to thermal treatment (annealing) under various conditions in order to obtain the samples with various oxygen contents [9]. As a result, three forms of the samples were obtained: "as grown," i.e., without annealing; "optimally reduced," i.e., optimally annealed in vacuum (60 min, $T$ = 780 $^0$C, $p$ = $10^{-2}$ mm of mercury); and "non optimally reduced," i.e., non optimally annealed in vacuum (40 min, $T$ = 780 $^0$C, $p$ = $10^{-2}$ mm of mercury). The thicknesses of the films were 1200–2000 Å. The temperature dependence of the resistance in the temperature range $T$ = 0.4–4.2 K in various magnetic fields up to 9 T was measured using the Oxford Instruments solenoid (Institute of Metal Physics, Ural Division, Russian Academy of Sciences).

**Experimental results and discussion**

Fig.1 shows the field dependence of the resistivity for optimally reduced $Nd_{1.85}Ce_{0.15}CuO_4$ single crystal film ($T_c$ = 21 K) at $T$ = 4.2 K. It is seen that $\rho_{xx}$ = 0 up to a field $B_p$ which is a vortex-depinning field or a field of melting of vortex lattice [10]. For $B > B_p$ resistivity $\rho_{xx}$ quickly increases with field until the normal state resistivity value $\rho_n$ is reached at the upper critical field $B_{c2}$. The upper critical field at given $T$ was extracted at the onset of superconducting transition defined by the intersection point from linear interpolation of $\rho_{xx}(B)$ at $B > B_p$ and $\rho_n(B)$ line as it is shown in Fig.1.

The interval $B_p < B < B_{c2}$ is the region of a mixed (vortex) state with the finite resistivity where dissipation in the system is a consequence of vortex moving under the action of the Lorentz force. When the pinning is included in the flux-flow model the resistivity $\rho_f$ for a sample in the mixed state for $\mathbf{J} \perp \mathbf{B}$ is given by [4, 11]:

$$\rho_f(B) = \rho_n \frac{B - B_p}{B_{c2}}, \tag{1}$$

so that dissipation is merely reduced by the fraction $(B-B_p)/B_{c2}$ as compared with a normal state.

In Fig.1 the field dependence of the Hall coefficient $R_H$ at $T$ = 4.2 K is also shown. The onset of nonzero Hall resistivity occurs at almost the same field (within experimental error) as does the onset of magnetoresistivity ($B_p$ = 3.2 T and $B_p$ = 3.5 T at $T$ =4.2 K respectively). The most remarkable feature of $R_H(B)$ dependence is a positive peak of $R_H$ in the mixed state $B_p < B \leq B_{c2}$ while in the normal state at $B > B_{c2}$ Hall coefficient $R_H$ is negative and essentially field independent.

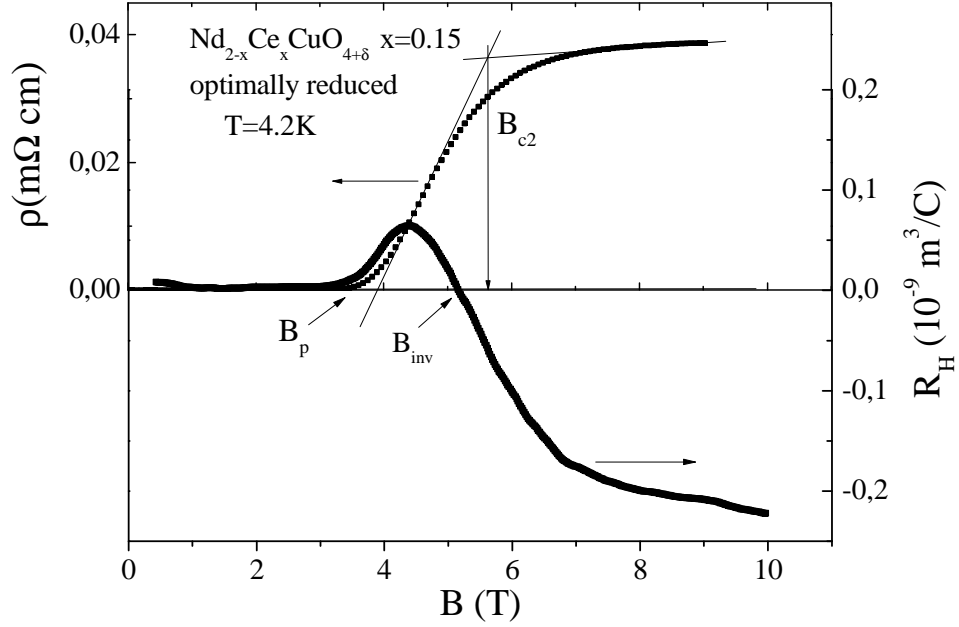

Fig.1. Field dependencies of the resistivity and Hall coefficient for optimally doped and optimally reduced $Nd_{1.85}Ce_{0.15}CuO_4$ at the temperature $T$=4.2 K.

In the treatment of [4, 11] which includes the back-flow current due to pinning forces, Hall resistivity for a mixed state is given by

$$\rho_{xy}^f = \rho_f(B)\mu B_{c2}\left[\left(1-\frac{B_p}{B}\right)-\left(1-\frac{B}{B_{c2}}\right)\frac{B_p}{B_{c2}}\right] \quad (2a)$$

for modified Nozieres-Vinen (NV) model and

$$\rho_{xy}^f = \rho_f(B)\mu B_{c2}\left(\frac{B}{B_{c2}} - \frac{2B_p}{B}\right) \quad (2b)$$

for modified Bardeen-Stephen (BS) model where $\mu$ is the charge carrier mobility. Then for the Hall coefficient $R_H^f = \rho_{xy}^f / B$ we have

$$R_H^f(B) = R_n \frac{B-B_p}{B}\left[\left(1-\frac{B_p}{B}\right)-\left(1-\frac{B}{2B_{c2}}\right)\frac{B_p}{B_{c2}}\right] \quad (3a)$$

and

$$R_H^f(B) = R_n \frac{B-B_p}{B}\left(\frac{B}{B_{c2}} - \frac{2B_p}{B}\right) \quad (3b)$$

in NV and BS approximations, respectively. Here $R_n = \rho_n \cdot \mu$ is the Hall coefficient in the normal state.

Equations (3a) and (3b) predict that at low fields the pinning terms will dominate over Lorentz terms, so that the Hall effect will have a sign opposite to that in the normal state. For optimally doped and optimally reduced sample for $T$= 4.2 K we really have a sign inversion of $R_H(B)$ at $B = B_{inv}$ = 5.2 T (see Fig.1). The estimations on formulas (3a) and (3b) with experimental values of parameters $B_p$ = 3.5 T and $B_{c2}$ =5.6 T give $B_{inv}^{NV}$ = 5.25 T and $B_{inv}^{BS}$ =6.3 T in very good accordance for NV model.

The evolution of $\rho_{xx}(B)$ and $R_H(B)$ dependencies with decrease of temperature for the same sample as in Fig.1 is shown in Fig.2. Both of $B_{c2}$ and $B_p$ values increase with decreasing of $T$ (see inset in Fig.2) and the region of the mixed state becomes narrower. Such a behavior is in accordance with existing physical conceptions of pinning forces: for instance, in models with normal vortex core interaction or magnetic interaction $B_p \sim B_{c2}^{3/2}$ [12].

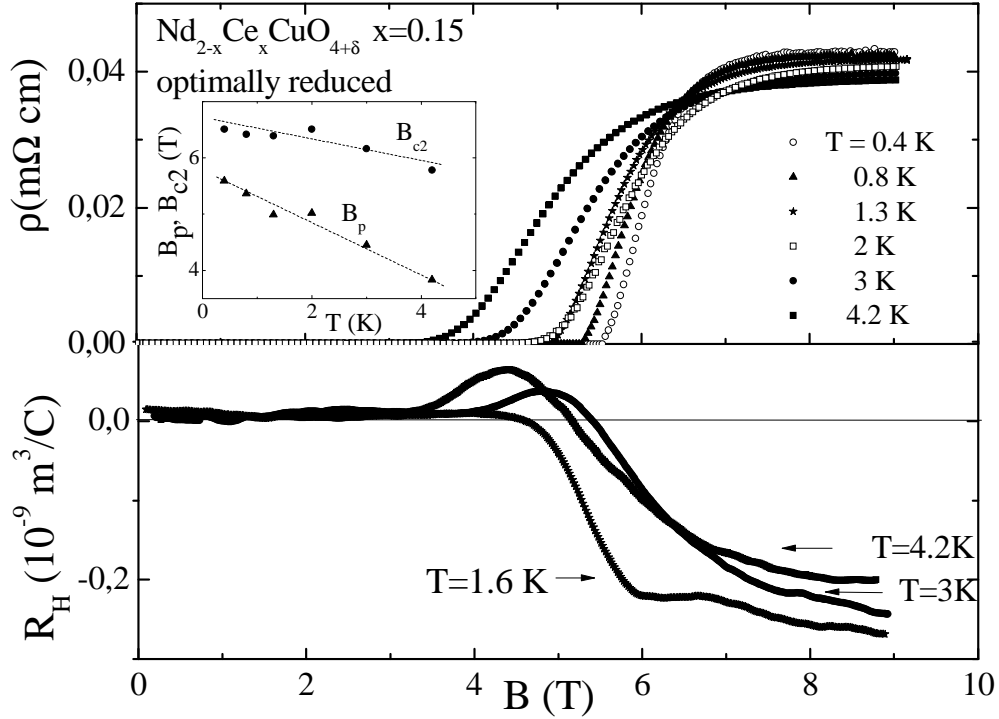

Fig.2. Field dependencies of the resistivity and Hall coefficient for optimally doped and optimally reduced $Nd_{1.85}Ce_{0.15}CuO_4$ at different temperatures.

The amplitude of the anomalous $R_H(B)$ peak depends on temperature, decreasing with its lowering in accordance with literature data [1, 7, 11] and vanishes for $T \leq 1.6$ K. In the normal state at $B > B_{c2}$ longitudinal resistivity is nearly independent on magnetic field and slightly (logarithmically) depends on temperature. The transition of Hall effect to the normal state is more or less completed at $B \geq 8$ T where $R_H$ becomes nearly constant. From the experimental values of the in-plane resistivity $\rho$ and Hall constant $R_H$ in the normal state, we have obtained the surface resistance $R_s = \rho/c$ per $CuO_2$ layer and the bulk and surface electron densities $n = (eR_H)^{-1}$ and $n_s = n \cdot c$ ($c = 6$ Å is the distance between $CuO_2$ layers). Using the relations $\sigma_s = (e^2/h)k_F\ell$ for the 2D conductance $\sigma_s = 1/R_s$, and $k_F = (2\pi n_s)^{1/2}$ for the Fermi wave vector, we have estimated the important parameter $k_F\ell$ and the mean free path $\ell$. These parameters for optimally reduced $Nd_{1.85}Ce_{0.15}CuO_4$ single crystal film are presented in the first line of the Table 1.

In Fig.3 we present magnetic field dependencies of the resistivity and Hall coefficient at $T = 4.2$ K for optimally doped $Nd_{2-x}Ce_xCuO_{4+\delta}$ single crystal films with different nonstoichiometric disorder. The parameter $(k_F\ell)^{-1}$ can serve as a measure of disorder in a random two-dimensional system ($CuO_2$ planes). It is seen that an increase of oxygen content leads to an increase of the degree of disorder (an increase of normal-state resistivity $\rho \sim (k_F\ell)^{-1}$) and a rapid decrease of the upper critical field.

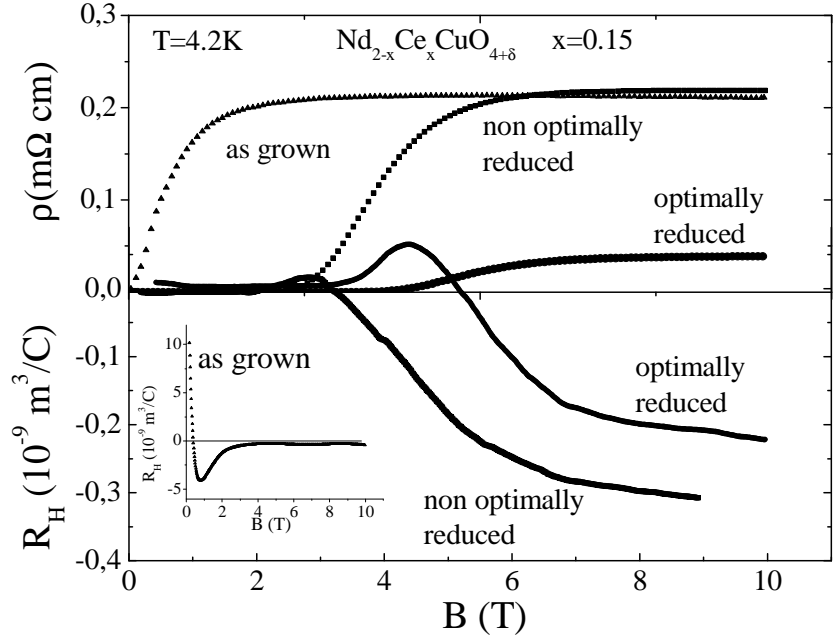

Fig.3. Field dependencies of the resistivity and Hall coefficient for optimally doped $Nd_{1.85}Ce_{0.15}CuO_{4+\delta}$ with different nonstoichiometric disorder at the temperature $T$=4.2 K.

We have found also that for optimally doped samples annealing in vacuum leads to an essential increase of the normal-state Hall coefficient (decrease of $|R_H|$) due to removing of the interstitial apical oxygen and delocalization of the charge carriers. After the next state of annealing in vacuum under optimal conditions a considerable increase of the mean free path of the carriers takes place (see Table 1).

As for the mixed state we see that the amplitude of anomalous Hall peak quickly drops with the increase of the degree of disorder (from optimally to non optimally reduced film) and for the most disordered film no sign reversal of the Hall effect in the mixed state is observed (see inset in Fig.3).

In Fig.4 we present field dependencies of the Hall coefficient at $T = 4.2$ K for optimally reduced ($\delta \rightarrow 0$) $Nd_{2-x}Ce_xCuO_4$ single crystals films with different cerium content. For underdoped ($x = 0.14$) and optimally doped ($x = 0.15$) samples at the normal state ($B \geq 6$ T for $x = 0.15$ and $B \geq 2.5$ T for $x = 0.14$) negative sign of $R_H$ is observed (that is the majority charge carriers are electrons) in accordance with the most of experimental data both for ceramics and single crystals [13-15]. It should be noted that the normal state Hall constant $|R_H|$ for $x = 0.14$ is an order of magnitude greater than $|R_H|$ for $x = 0.15$ and thus at low temperatures the electron density in underdoped region is an order of magnitude lower as compared with optimally doped region (Table 2).

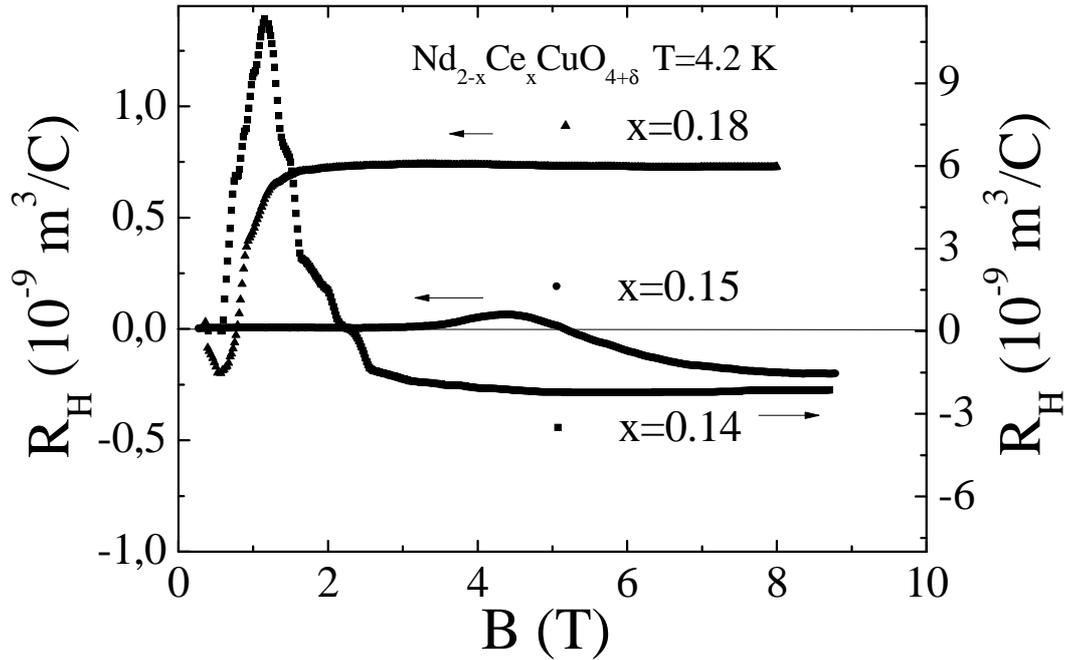

Fig.4. Field dependencies of the resistivity and Hall coefficient for optimally reduced $Nd_{2-x}Ce_xCuO_{4+\delta}$ at the temperature $T=4.2$ K.

Table 1. Parameters in the normal state of optimally doped ($x = 0.15$) $Nd_{2-x}Ce_xCuO_{4+\delta}$.

| № * | $\rho \cdot 10^3 [\Omega cm]$ | $k_F \ell$ | $R_H \times 10^3$ [sm$^3$/C] | $n \times 10^{-22}$ [sm$^{-3}$] | $n_s \times 10^{-14}$ [sm$^{-2}$] | $k_F \times 10^{-7}$ [sm$^{-1}$] | $\ell$ [Å] |
|---|---|---|---|---|---|---|---|
| 1 | 0.03 | 51.6 | -0.23 | 2.72 | 16.3 | 10.1 | 51.3 |
| 2 | 0.15 | 9.1 | -0.35 | 1.81 | 10,86 | 8.3 | 12.5 |
| 3 | 0.17 | 8.6 | -0.92 | 0.68 | 4.1 | 5.1 | 18.3 |

*(1 – optimally reduced, 2- non optimally reduced, 3 – as grown.)

Table 2. Parameters in the normal state of optimally reduced $Nd_{2-x}Ce_xCuO_{4+\delta}$.

| Samples | $T_c$ [K] | $\rho \times 10^3 [\Omega \cdot sm]$ | $k_F \ell$ | $R_H \times 10^3$ [sm$^3$/C] | $n \times 10^{-22}$ [sm$^{-3}$] | $k_F \cdot 10^{-7}$ [sm$^{-1}$] | $\ell$ [Å] |
|---|---|---|---|---|---|---|---|
| $x = 0.14$ | 11 | 0.59 | 2.7 | -2.3 | 0.27 | 3.2 | 8.2 |
| $x = 0.15$ | 21 | 0.03 | 51.6 | -0.23 | 2.72 | 10.1 | 51.3 |
| $x = 0.18$ | 6 | 0.04 | 44.4 | +0.73 | 0.86 | 5.7 | 78.0 |

Positive sign of the Hall effect is firmly observed in the normal state of overdoped sample ($x = 0.18$) as well as for $x \geq 0.17$ in the normal state at $T > T_c$ ([16] and references therein).

As for the mixed state, it is remarkable that at $T = 4.2$ K the Hall anomaly (the sign change of the Hall effect) is clearly seen both for $R_H < 0$ ($x = 0.14$ and $x = 0.15$) and for $R_H > 0$ ($x = 0.18$) (see Fig.4). Thus, the essential behavior of anomalous Hall effect is independent of the sign of the majority charge carriers that as argued by Hagen et al. [5] supports a vortex dynamics interpretation.

Although NV and BS models with finite pinning force could in principle explain the systematic change of Hall effect sign in a mixed state, in opinion of many authors (see, for instance, [5,11, 17, 18] more theoretical work is needed to complete the understanding of vortex motion in type-II superconductors. Thus, a novel mechanism for the sign change of the Hall effect in the flux flow region is proposed by Feigelman et al. [17]. They show that the sign change may follow from the difference $\delta n$ between the electron density at the center of the vortex core and the density for outside the core.

**Conclusions**

So, the behaviors of the Hall coefficient and magnetoresistivity in the mixed state of electron doped superconductor $Nd_{2-x}Ce_xCuO_{4+\delta}$ with different Ce content ($x$) and nonstoichiometric disorder ($\delta$) have been discussed in the framework of the flux-flow NV and BS models with back flow effect due pinning forces taken into account.

We also investigate an effect of nonstoichiometric disorder on the resistivity and Hall effect in the normal state of optimally doped superconducting $Nd_{1.85}Ce_{0.15}CuO_{4+\delta}$ single crystal films.

For the optimally doped $Nd_{2-x}Ce_xCuO_{4+\delta}$ (x=0.15) the heat treatment (annealing) under various conditions leads to the decrease of the disorder parameter, charge carriers delocalization and considerable increase of the mean free path.

**Acknowledgements**

This work was done within RAS Program (project N 01.2.006.13394) and with partial support of RFBR N 09-02-96518.